\documentclass[doublecol]{epl2} 
\usepackage{amsmath,amssymb}

\title{Nonlinear adiabatic response of interacting quantum dots}

\author{Oleksiy Kashuba\thanks{E-mail: \email{kashuba@physik.rwth-aachen.de}} 
\and Herbert Schoeller \and Janine Splettstoesser}
\shortauthor{O.~Kashuba, H.~Schoeller and J.~Splettstoesser}
\institute{Institut f\"ur Theorie der Statistischen Physik, RWTH Aachen, 52056 Aachen, 
Germany, EU and JARA - Fundamentals of Future Information Technology}


\abstract
{We develop a generic method in Liouville space to describe the dissipative dynamics of 
coherent interacting quantum dots with adiabatic time dependence beyond linear response.
We show how the adiabatic response can be related to effective quantities known
from real-time renormalization group methods for stationary quantities.
We propose the study of a delay time as a characteristic time scale. We apply the 
method to the interacting resonant level model and calculate the nonlinear adiabatic 
charge response to time-dependent gate voltages, tunneling barriers and Coulomb interaction. 
The dot charge  delay time is found to be given by a unique 
expression in all cases, in contrast to the capacitance and the charge relaxation 
resistance. We discuss the observability of the effects in molecular systems and cold-atom 
setups.}

\date{\today}

\newsavebox{\dlgamma}
\savebox{\dlgamma}(40,12){%
\thicklines
\put(-17.6,12){\color{green}\linethickness{1.3pt}\line(1,0){35.2}}
\put(-17,3){\color{green}\linethickness{1.3pt}\line(0,1){9.6}}
\put(17,3){\color{green}\linethickness{1.3pt}\line(0,1){9.6}}
\put(-14,0){\linethickness{1.3pt}\line(1,0){28}}
\put(-17,0){\circle{6}}
\put(17,0){\circle{6}}
}
\newsavebox{\dluu}
\savebox{\dluu}(46,12){%
\thicklines
\put(-20.6,12){\color{green}\linethickness{1.3pt}\line(1,0){41.2}}
\put(-14.6,6){\color{green}\linethickness{1.3pt}\line(1,0){29.2}}
\put(-20,3){\color{green}\linethickness{1.3pt}\line(0,1){9.6}}
\put(20,3){\color{green}\linethickness{1.3pt}\line(0,1){9.6}}
\put(-14,3){\color{green}\linethickness{1.3pt}\line(0,1){3.6}}
\put(14,3){\color{green}\linethickness{1.3pt}\line(0,1){3.6}}
\put(-11,0){\linethickness{1.3pt}\line(1,0){22}}
\put(-20,0){\circle{6}}
\put(-14,0){\circle{6}}
\put(20,0){\circle{6}}
\put(14,0){\circle{6}}
}                           

\begin{document}

\maketitle

\section{Introduction}
Adiabatic transport through quantum dots associated with a slow cyclic time dependence of 
the system parameters has generated a lot of interest in recent years, particularly in 
connection with quantum pumps~\cite{buttiker0,brouwer,pothier,giazzotto} and mesoscopic 
capacitors~\cite{buttiker1,gabelli}, see also Ref.~\cite{review_nt}.
For noninteracting systems the scattering formalism is a powerful tool to describe the 
adiabatic response~\cite{buttiker0,brouwer}.
A central challenge in this field is the understanding of the influence of strong 
interactions as they typically occur in small quantum dots.
Although general current formulas have been derived in terms of Green's 
functions~\cite{splett0,selaoreg,fiorettosilva}, their evaluation is quite difficult 
in the coherent regime at low temperature.
Progress has been achieved in the perturbative regime of high temperature~\cite{splett1}, 
where it was shown that pure interaction effects can be revealed by the adiabatic response 
which would be covered by more dominant effects in the steady state~\cite{splett1,reckermann10}.
These studies also included the properties of the $RC$-time in linear response~\cite{splett2}.
In contrast, for interacting quantum dots at low temperature, a generic formalism 
providing the adiabatic time evolution in response to any parameter in linear or nonlinear response
is not yet available. So far, quantum pumping has been studied for special models, like e.g.~the 2-channel 
Kondo model in the strong coupling regime~\cite{selaoreg}, the Kondo model at the exactly 
solvable Toulouse point~\cite{schiller2}, and the single-impurity Anderson model within 
slave-boson mean-field approximation~\cite{aono}. In addition, the research on interaction effects
in mesoscopic capacitors has concentrated on the special case of linear charge response to 
an external AC gate voltage by using the standard relation to the equilibrium density-density 
correlation function. Here, the main 
object of interest was the charge relaxation resistance $R$ defined by expanding the
charge response $\delta Q$ in the external frequency $\omega$ via
\begin{equation}
\delta Q = ( C + i\omega RC^{2}) \delta \mathcal{V} \quad,
\label{eq:RC}
\end{equation}
where $Q(t)=\delta Q e^{-i\omega t}$ denotes the charge, $C$ is the static quantum capacitance 
and $\mathcal{V}(t)=\delta \mathcal{V}e^{-i\omega t}$ defines the external AC gate voltage.
For a single transport channel and provided that the Coulomb interaction is weak, a universal relaxation resistance $R=h/2e^2$ was found in 
the coherent regime at zero temperature~\cite{buttiker1,buttiker2}.
For interacting metallic dots and the single-impurity Anderson 
model the Shiba relation was shown to be a powerful tool to analyze $R$ and its 
universality~\cite{rodionov,lehur-ondotint,lehur_preprint}.
Using bosonization, the influence of Luttinger-liquid leads on $R$ has also been 
studied~\cite{kato-inleadint} and a numerical approach has been used away from the 
Fermi-liquid regime~\cite{leelopez}. 

In this Letter we develop a general approach to deal with the adiabatic dissipative 
response, where the time scale of the external modulation
$\propto\omega^{-1}$ is much larger than the inverse of typical relaxation rates 
$\Gamma_{c}$, of a coherent quantum dot at low temperature including  Coulomb interactions. 
We show that the adiabatic response can be calculated very efficiently
by using quantum field theoretical methods in Liouville space developed in 
Refs.~\cite{hs_epj09,leijnse_wegewijs_prb08} and generalized here for the case of 
time-dependent Hamiltonians (for approaches within Keldysh formalism
see the review articles \cite{review_nt,hs_epj09} and the recent development \cite{kennes_meden}).
We provide a general relationship of the adiabatic response to effective Liouvillians and 
vertices known from real-time renormalization group (RTRG) in the stationary limit with 
instantaneous time parametrization, based on powerful techniques for the calculation of 
Laplace-variable derivatives, recently used within the $E$-flow 
scheme of  RTRG~\cite{pletyukhov_hs_preprint}.
As a consequence, our formalism is suitable to any model which can be treated by RTRG,
which is applicable for many generic models with charge and 
spin fluctuations~\cite{hs_epj09}. Recent applications of RTRG cover the Kondo model for both weak 
\cite{hs_reininghaus_prb09,pletyukhov_schuricht_hs_prl10,schuricht_hs_prb09,
pletyukhov_schuricht_prb11,hoerig_etal_prb12} and strong~\cite{pletyukhov_hs_preprint} coupling,
and the interacting resonant level model (IRLM)~\cite{rtrg_irlm}. 
Most importantly, in contrast to previous research, our formalism allows for
the adiabatic variation of {\it any} parameter in {\it nonlinear} response, 
where no relation to equilibrium density-density correlation functions is possible and
where certain identities like e.g.~the Shiba relation are no longer applicable.
Therefore, going beyond previous studies of the linear response to an external gate voltage, 
we also study the response to other parameters, like 
the tunneling coupling or the Coulomb interaction, which experimentally can be either
realized intentionally, or indirectly induced by the gate voltage.
We even cover the regime of nonlinear response, motivated by recent works on mesoscopic 
capacitors in the nonlinear driving regime~\cite{nonlinear}.
Instead of the linear response formula~(\ref{eq:RC}) for the charge variation by an external 
gate voltage, we decompose the dynamics of any observable $A$, with a nonvanishing 
instantaneous contribution and its adiabatic correction in response to any parameter, as 
$A(t)=A^{(i)}(t)+A^{(a)}(t)$.
The central quantity of our interest is the delay time scale $\tau_A$ for the expectation 
value $A$, defined by
\begin{equation}
\label{eq:delay_time}
\tau_A\,=\, |A^{(a)}/\dot{A}^{(i)}|\quad,
\end{equation}  
which describes the delay of the full solution comparing to the instantaneous one.
For $A\equiv \dot{Q}$ in linear response to a time-dependent gate voltage, it is equivalent 
to the $RC$-time. In general $\tau_A$ can be quite different from typical relaxation times, 
depending on the observable, the type of excitation and its amplitude, and it is of 
fundamental interest to understand its dependence on interactions. 

We use our method to consider the IRLM with a single lead, which constitutes a 
minimal model for the mesoscopic capacitor 
with one single-particle level, where strong correlations are induced by a local Coulomb 
interaction between the dot and the lead.
Recently, the IRLM has been extensively used to study nonequilibrium transport through 
interacting quantum dots~\cite{mehta,saleur1,saleur2,rtrg_irlm,karrasch10}, including 
the dynamics of the time evolution into the stationary state~\cite{rtrg_irlm}.
We calculate the nonlinear adiabatic charge response and the delay time $\tau_Q$, 
including renormalization effects of the tunneling enhanced by correlations.
Importantly, we find that the functional form of the charge delay time $\tau_Q$ is
robust against the choice of the time-dependent parameter even in nonlinear response, 
whereas the capacitance $C$ or the relaxation resistance $R$ get a more complex form when 
the tunneling or the Coulomb interaction are varied.
Finally, we analyze further possible experimental implementations of the predicted results 
for the IRLM with time-dependent parameters, namely via the Anderson-Holstein model in 
molecular electronics or via the spin-boson model in cold-atom setups.

\section{Method}
We start from a general Hamiltonian 
$H(t)=H_{\rm res}+\sum_\alpha\mu_\alpha(t)\hat{N}_\alpha+H_{\rm dot}(t)+V(t)$ of an interacting 
quantum dot coupled to noninteracting fermionic reservoirs with time-dependent chemical potentials
$\mu_\alpha(t)$ and a flat d.o.s. of width $2D$ via the coupling $V(t)$.
Generalizing the Liouvillian approach of Ref.~\cite{hs_epj09} to the case of 
time-dependent Hamiltonians, one finds that the dissipative dynamics of the reduced density 
matrix $\rho(t)$ of the dot can be described by the effective Liouvillian equation
\begin{equation}
i\partial_{t}\rho(t) = \int_{t_{0}}^{t} L(t,t') \rho(t') dt',
\label{eq:rhoeq}
\end{equation}
where $L(t,t')$ is the effective dot Liouvillian obtained by integrating out the reservoirs ($\hbar=1$). 
At the initial time $t_{0}$ the total density matrix factorizes into an 
arbitrary dot and an equilibrium reservoir part. Since we are only interested in the 
asymptotic dynamics we set $t_0=-\infty$ below.
Following Ref.~\cite{hs_epj09}, the effective Liouvillian can be calculated 
diagrammatically, where each diagram of order $O(V^n)$ consists of a product of 
vertices $G(t_i)$, $t_1>\dots >t_n$, with effective dot propagators $\Pi(t_i,t_{i-1})$ in between.
In addition, the information of the Fermi distribution and the d.o.s.~of the reservoirs 
is contained in time-independent reservoir contractions between the vertices.
Using the formal definition $G(t,t')=G(t)\delta(t-t'-0^+)$, we find that each term can 
be written in terms of a generalized convolution in time space as 
$(G\circ\Pi\circ\dots\circ\Pi\circ G)(t,t')$, where 
$(A\circ B)(t,t')\equiv \int_{t'}^{t} d\tau A(t,\tau)B(\tau,t')$.
Introducing the partial Laplace transform $A(t;E)=\int_{-\infty}^{t} dt' e^{i(E+i0)(t-t')}A(t,t')$,
we get 
\begin{align}
\nonumber
&(A_1\circ A_2\circ \dots \circ A_n)(t;E) =e^\mathcal{D} A_1(t;E)\dots A_n(t;E)=\\
\label{eq:gradient_expansion}
&\left.e^{i\sum_{j>k}\partial_{E_{j}}\partial_{t_{k}}} A_1(t_{1},E_{1})
\dots A_n(t_{n},E_{n})\right|_{\scriptsize E_{j}=E,t_{k}=t}\,.
\end{align}
The special differential operator $\mathcal{D}=i\partial_E^{\text{left}}\partial_t^{\text{right}}$ prescribes the 
energy derivative to act left to the time derivative.
This rule is a natural generalization to Laplace space of analog identities in Fourier 
space used for gradient expansions in the Keldysh formalism~\cite{rammer}.
Formally, it allows for the straightforward application of the Liouvillian approach to time-dependent 
Hamiltonians, with the difference that the exponential differential operator has to be 
taken beforehand. In the adiabatic case, the exponential can be expanded in 
$\partial_E\partial_t\sim \frac{\omega}{\Gamma_{c}}\ll 1$, leading to an expansion of the 
effective Liouvillian,
\begin{equation}
\label{eq:L_adiabatic_expansion}
L(t;E) \,=\, L^{(i)}(t;E) \,+\, L^{(a)}(t;E) \,+\,\dots\quad.
\end{equation}
Here, $L^{(i)}(t;E)$ denotes the instantaneous part, where the time $t$ enters only 
parametrically via the external parameters, and $L^{(a)}(t;E)$ is the first adiabatic 
correction, which is linear in the time derivatives of the external parameters.

Once the effective Liouvillian $L(t;E)$ is known up to the adiabatic correction, one can use 
it in eq.~(\ref{eq:rhoeq}) which reads 
\begin{equation}
i\partial_{t}\rho(t) = 
(L\circ\rho)(t;0) = \left.e^{i \partial^{L}_{E} \partial^{\rho}_{t}} L(t;E) \rho(t)\right|_{E=0}
\label{eq:tE_kinetic_equation}
\end{equation}
in the mixed $(t;E)$-representation. Expanding $\rho(t)=\rho^{(i)}(t)+\rho^{(a)}(t)+\dots$ analogously to 
eq.~(\ref{eq:L_adiabatic_expansion}), we find by comparing equal powers in the external frequency 
\begin{align}
\label{eq:inst}
L^{(i)}\rho^{(i)}=0 \quad,\quad \mathrm{Tr}\,\rho^{(i)}=1 \quad,\quad \mathrm{Tr}\,\rho^{(a)}=0 \quad,
\\
\label{eq:adia}
L^{(i)}\rho^{(a)} + L^{(a)}\rho^{(i)} - i(1-\partial_{E}L^{(i)})\partial_{t}\rho^{(i)} =0 \quad .
\end{align}
In all arguments of $L^{(i/a)}$ and $\partial_{E}L^{(i)}$, $E=0$ has to be taken.
From these equations the instantaneous density matrix $\rho^{(i)}(t)$ and the first 
adiabatic correction $\rho^{(a)}(t)$ can be determined.
We emphasize that this approach is even applicable in nonlinear response in the amplitude 
of the external perturbations, i.e.~only the  time scale of the external modulation needs 
to be large enough.
Furthermore, it allows for an adiabatic modulation of any parameter of the Hamiltonian and 
is not restricted to a time-dependent gate voltage.

The algebra of~(\ref{eq:inst}) and~(\ref{eq:adia}) can 
be easily evaluated for quantum dots with two accessible states. If additional conservation laws 
are present (as, \textit{e.g.}, charge conservation
in the IRLM or spin-$S_z$ conservation in the Kondo model), the nonvanishing matrix elements 
of the Liouvillian can be written as
\begin{align}
\label{eq:Liouvillian}
L_{\bar{s}\bar{s},ss}=-L_{ss,ss}=i\Gamma_{s}=i\Gamma/2+is\Gamma'
\,\,,\,\,
L_{s\bar{s},s\bar{s}}=\epsilon_s\,,
\end{align}
where $s\equiv\pm$ denotes the two states and $\bar{s}=-s$. 
At $E=0$ we get from~(\ref{eq:inst}) and~(\ref{eq:adia}) that the instantaneous density matrix 
is diagonal $\rho^{(i)}_s=\Gamma^{(i)}_s/\Gamma^{(i)}=
1/2+s \Gamma^{\prime(i)}/\Gamma^{(i)}$ and the
adiabatic correction fulfills $\rho^{(a)}_+=-\rho^{(a)}_-$ with
\begin{align}
\label{eq:rho_adiab}
\rho^{(a)}_+=\frac{1}{\Gamma^{(i)}}\left\{\Gamma^{\prime (a)}-
\frac{\Gamma^{\prime(i)}}{\Gamma^{(i)}}\Gamma^{(a)}-
(1+i\partial_E\Gamma^{(i)})\partial_t\left(\frac{\Gamma^{\prime(i)}}{\Gamma^{(i)}}\right)\right\}
\end{align}
Below we use this result to evaluate the adiabatic response for the IRLM.

\section{Calculation of $L^{(a)}$}
We now turn to the central issue of how to relate the adiabatic correction 
$L^{(a)}(t;E)$ to the instantaneous quantities known from RTRG.
Using~(\ref{eq:gradient_expansion}), we can formally write 
$L^{(a)}(t;E)=\mathcal{D} L^{(i)}(t;E)=i\partial_E^{\text{left}}\partial_t^{\text{right}} L^{(i)}(t;E)$. The representation of $L^{(i)}(t;E)$ by its diagrammatic expansion specifies
what ``left'' and ``right'' means for the derivatives with respect to $E$ and $t$.
As a first step, we represent the derivative $\partial_E L^{(i)}$ by effective vertices and
propagators, using a method developed in Ref.~\cite{pletyukhov_hs_preprint}.
For generic models with two types of vertices, namely single (e.g. tunneling) and double (e.g. Coulomb interaction, exchange, etc.), 
we decompose it into two contributions in leading order and find
\begin{align}
\nonumber
\partial_E L^{(i)}(t;E)&=\quad\partial_E L^{(i)}_\Gamma(t;E)+\partial_E L^{(i)}_U(t;E)\\ 
&= \,
\begin{picture}(50,15)(-25,0)
\put(0,0){\usebox{\dlgamma}}
\thicklines
\color{red}
\qbezier(-3,-2.5)(0,0)(3,2.5)
\end{picture}
+
\frac{1}{2}
\begin{picture}(50,15)(-25,0)
\put(0,0){\usebox{\dluu}}
\thicklines
\color{red}
\qbezier(-3,-2.5)(0,0)(3,2.5)
\end{picture}\quad.
\label{eq:L_E_der}
\end{align}
The diagrammatic rules are explained in detail in Refs.~\cite{hs_epj09,pletyukhov_hs_preprint}.
The single (double) circles represent full effective single (double) vertices with effective propagators 
$\Pi^{(i)}(t;E)=\frac{1}{E-L^{(i)}(t;E)}$ in between (the Laplace variable is shifted by the frequencies
and chemical potentials of all reservoir contractions crossing over the propagator). 
The left slash indicates $\partial_E$ and the grey (green, color online) line 
represents the reservoir contraction given by the antisymmetric part $f(\omega)-\frac{1}{2}$ of
the Fermi distribution function. All possible diagrams for $\partial_{E}L^{(i)}$ can be classified 
by the number of lines over the 
propagator containing a derivative. In the next step we perform the time derivative $i\partial_t$
right to the energy derivative. The energy derivative is then shifted by partial integration to
the reservoir contraction (indicated by a (blue) cross)~\cite{pletyukhov_hs_preprint}. This yields
\begin{align}
\nonumber
L^{(a)}_\Gamma(t;E)\quad &= \quad\quad\!
\begin{picture}(50,15)(-25,0)
\put(0,0){\usebox{\dlgamma}}
\thicklines
\color{red}
\qbezier(-3,-2.5)(0,0)(3,2.5)
\qbezier(19.5,-2.5)(17,0)(14.5,2.5)
\end{picture}
+
\begin{picture}(50,15)(-25,0)
\put(0,0){\usebox{\dlgamma}}
\thicklines
\color{red}
\qbezier(-4,-2.5)(-2,0)(0,3)
\qbezier(0,3)(2,0)(4,-2.5)
\end{picture}
\\
\label{eq:LG_Et_der}
&= \quad
-
\begin{picture}(50,15)(-25,0)
\put(0,0){\usebox{\dlgamma}}
\thicklines
\color{blue}
\qbezier(-2.5,9.5)(0,12)(2.5,14.5)
\qbezier(-2.5,14.5)(0,12)(2.5,9.5)
\color{red}
\qbezier(19.5,-2.5)(17,0)(14.5,2.5)
\end{picture}
+
\begin{picture}(50,15)(-25,0)
\put(0,0){\usebox{\dlgamma}}
\thicklines
\color{red}
\qbezier(-4,-2.5)(-2,0)(0,3)
\qbezier(0,3)(2,0)(4,-2.5)
\end{picture}\quad,
\\
\label{eq:LU_Et_der}
L^{(a)}_U(t;E)\quad &= \quad \frac{1}{2}
\begin{picture}(50,15)(-25,0)
\put(0,0){\usebox{\dluu}}
\thicklines
\color{red}
\qbezier(-3,-2.5)(0,0)(3,2.5)
\qbezier(23,-2.5)(17,0)(11,2.5)
\end{picture}
+\,\,\frac{1}{2}
\begin{picture}(50,15)(-25,0)
\put(0,0){\usebox{\dluu}}
\thicklines
\color{red}
\qbezier(-4,-2.5)(-2,0)(0,3)
\qbezier(0,3)(2,0)(4,-2.5)
\end{picture}\quad,
\end{align}
where the right slash represents $i\partial_t$ and the hat indicates the differential
operator $\mathcal{D}=i\partial_E^{\text{left}}\partial_t^{\text{right}}$.
The frequency integral
in both diagrams of~(\ref{eq:LG_Et_der}) is well-defined in the wide-band limit, so 
(\ref{eq:LG_Et_der}) provides an explicit expression for the adiabatic 
correction containing the
tunneling vertices in terms of renormalized vertices and propagators. In contrast,
the frequency integrals in~(\ref{eq:LU_Et_der}) are logarithmically
divergent. We therefore take a second derivative with respect to $E$, yielding an RG equation 
for the adiabatic part, $L^{(a)}_U(t;E)$, 
after partial integration. This contains the
double vertices
\begin{align}
\label{eq:LU_EEt_der}
\partial_E\,L^{(a)}_U(t;E)\, &= \, \frac{1}{2}
\begin{picture}(50,15)(-25,0)
\put(0,0){\usebox{\dluu}}
\thicklines
\color{blue}
\qbezier(-2.5,9.5)(0,12)(2.5,14.5)
\qbezier(-2.5,14.5)(0,12)(2.5,9.5)
\qbezier(-2.5,3.5)(0,6)(2.5,8.5)
\qbezier(-2.5,8.5)(0,6)(2.5,3.5)
\color{red}
\qbezier(23,-2.5)(17,0)(11,2.5)
\end{picture}
-\,\,\frac{1}{2}
\begin{picture}(50,15)(-25,0)
\put(0,0){\usebox{\dluu}}
\thicklines
\color{blue}
\qbezier(-2.5,9.5)(0,12)(2.5,14.5)
\qbezier(-2.5,14.5)(0,12)(2.5,9.5)
\color{red}
\qbezier(-4,-2.5)(-2,0)(0,3)
\qbezier(0,3)(2,0)(4,-2.5)
\end{picture}\quad.
\end{align}
Eqs.~(\ref{eq:LG_Et_der}) and~(\ref{eq:LU_EEt_der}) are the final results for the evaluation of 
adiabatic corrections of the Liouvillian in leading order, based on the 
instantaneous values of the renormalized vertices and Liouvillian, which are obtained
from RTRG\footnote{%
Provided that the frequency integrals converge, we note that our results 
can even be applied to a frequency-dependent d.o.s. in the leads. Otherwise,
another derivative with respect to $E$ may be required.}.
For the adiabatic part of the propagator, appearing in the second 
diagram of~(\ref{eq:LG_Et_der}) and~(\ref{eq:LU_EEt_der}) each, we insert
$\Pi^{(a)}=\Pi^{(i)} L^{(a)}\Pi^{(i)} + 
(\partial_E\Pi^{(i)})(i\partial_t L^{(i)})\Pi^{(i)}$. The first term 
does however not contribute to the adiabatic propagator in leading order.

An interesting question is whether derivatives with respect to
the Laplace and time variable commute in leading order, i.e. whether the
adiabatic correction to the effective Liouvillian, eqs.~(\ref{eq:LG_Et_der}) and~(\ref{eq:LU_EEt_der}), 
can be written as 
\begin{equation}
L^{(a)}_{\Gamma/U}(t;E) \stackrel{?}{=} \frac{1}{2}i\partial_{E}\partial_{t}L^{(i)}_{\Gamma/U}(t;E)\, .
\label{eq:L_a_approx}
\end{equation}
A similar relation was investigated so far only for noninteracting systems~\cite{moskalets}.
To analyze its validity we introduce the complementing differential operator
$\mathcal{D}'=i\partial_E^{\text{right}}\partial_t^{\text{left}}$, where the energy derivative
is taken {\it right} to the time derivative. Analogously to~(\ref{eq:LG_Et_der}) and~(\ref{eq:LU_EEt_der}) one finds in leading order 
\begin{align}
\label{eq:LG_Et_der_c}
\mathcal{D}'L^{(i)}_\Gamma(t;E)\,\, &= \,\, -
\begin{picture}(50,15)(-25,0)
\put(0,0){\usebox{\dlgamma}}
\thicklines
\color{blue}
\qbezier(-2.5,9.5)(0,12)(2.5,14.5)
\qbezier(-2.5,14.5)(0,12)(2.5,9.5)
\color{red}
\qbezier(-14.5,-2.5)(-17,0)(-19.5,2.5)
\end{picture}
+
\begin{picture}(50,15)(-25,0)
\put(0,0){\usebox{\dlgamma}}
\thicklines
\color{red}
\qbezier(0,-2.5)(-2,0)(-4,3)
\qbezier(4,3)(2,0)(0,-2.5)
\end{picture}\quad,
\\
\label{eq:LU_EEt_der_c}
\partial_E\,\mathcal{D}'L^{(i)}_U(t;E)\, &= \, \frac{1}{2}
\begin{picture}(50,15)(-25,0)
\put(0,0){\usebox{\dluu}}
\thicklines
\color{blue}
\qbezier(-2.5,9.5)(0,12)(2.5,14.5)
\qbezier(-2.5,14.5)(0,12)(2.5,9.5)
\qbezier(-2.5,3.5)(0,6)(2.5,8.5)
\qbezier(-2.5,8.5)(0,6)(2.5,3.5)
\color{red}
\qbezier(-11,-2.5)(-17,0)(-23,2.5)
\end{picture}
-\,\,\frac{1}{2}
\begin{picture}(50,15)(-25,0)
\put(0,0){\usebox{\dluu}}
\thicklines
\color{blue}
\qbezier(-2.5,9.5)(0,12)(2.5,14.5)
\qbezier(-2.5,14.5)(0,12)(2.5,9.5)
\color{red}
\qbezier(0,-2.5)(-2,0)(-4,3)
\qbezier(4,3)(2,0)(0,-2.5)
\end{picture}\,\,.
\end{align}
The inverted hat represents the differential operator $\mathcal{D}'$.
Using $i\partial_E\partial_t = \mathcal{D}+\mathcal{D}'$, we can write
$\mathcal{D}=\frac{1}{2}i\partial_E\partial_t + \frac{1}{2}(\mathcal{D}-\mathcal{D}')$ and, thus,
the correction to eq.~(\ref{eq:L_a_approx}) for $L^{(a)}_{\Gamma}$ 
($\partial_E L^{(a)}_U$) is given by half the difference
of~(\ref{eq:LG_Et_der}) and~(\ref{eq:LG_Et_der_c}) ((\ref{eq:LU_EEt_der})
and~(\ref{eq:LU_EEt_der_c})). We first address the second diagrams on the r.h.s. of 
these equations: their differences involve the expression
\begin{align}
\label{eq:prop}
\frac{1}{2}(\mathcal{D}-\mathcal{D}')\Pi^{(i)}&=\Pi^{(i)}\left(\frac{1}{2}(\mathcal{D}-\mathcal{D}')L^{(i)}\right)\Pi^{(i)}\\
\nonumber
&\hspace{-2cm}
+\frac{1}{2}\left\{(\partial_E\Pi^{(i)})(i\partial_t L^{(i)})\Pi^{(i)}-
\Pi^{(i)}(i\partial_t L^{(i)})(\partial_E \Pi^{(i)})\right\}
\end{align}
for the propagator. Here, the first term on the r.h.s. can be neglected in leading
order, whereas the second one is only zero if the Liouvillian and 
its time and energy derivative commute. For special cases this is indeed possible:
it follows trivially for blocks where the Liouvillian is diagonal, as e.g.
the $2\times 2$-block $L^{(i)}_{s\bar{s},s'\bar{s}'}=\delta_{ss'}\epsilon^{(i)}_s$ of eq.~(\ref{eq:Liouvillian}). For $2$-level systems with conservation
laws, see eq.~(\ref{eq:Liouvillian}), it holds
also for the block $L^{(i)}_{ss,s's'}$ since the zero eigenvalue of
the Liouvillian can be omitted in a propagator standing left to a vertex averaged over
the Keldysh indices~\cite{hs_epj09}. Therefore, for this block one can 
replace the Liouvillian by its nonzero eigenvalue $-i\Gamma^{(i)}(t;E)$ and the
second term on the r.h.s.~of~(\ref{eq:prop}) is again zero. If this is given 
(or if the term can be neglected in leading order for certain models), we can write the 
correction to eq.~(\ref{eq:L_a_approx}) generically as
\begin{align}
\nonumber
L^{(a)}_\Gamma(t;E)\,\, &= \,\,\frac{1}{2}i\partial_E\partial_t L^{(i)}_\Gamma(t;E) 
\\ \label{eq:LG_final}
&+\,\,\frac{1}{2}
\begin{picture}(50,15)(-25,0)
\put(0,0){\usebox{\dlgamma}}
\thicklines
\color{blue}
\qbezier(-2.5,9.5)(0,12)(2.5,14.5)
\qbezier(-2.5,14.5)(0,12)(2.5,9.5)
\color{red}
\qbezier(-14.5,-2.5)(-17,0)(-19.5,2.5)
\end{picture}
-\,\,\frac{1}{2}
\begin{picture}(50,15)(-25,0)
\put(0,0){\usebox{\dlgamma}}
\thicklines
\color{blue}
\qbezier(-2.5,9.5)(0,12)(2.5,14.5)
\qbezier(-2.5,14.5)(0,12)(2.5,9.5)
\color{red}
\qbezier(19.5,-2.5)(17,0)(14.5,2.5)
\end{picture}\quad,
\\ 
\nonumber
\partial_E L^{(a)}_U(t;E)\,\, &= \,\,
\partial_E\left\{\frac{1}{2}i\partial_E\partial_t L^{(i)}_U(t;E) \right\}
\\ 
\label{eq:LU_final}
&+\,\,\frac{1}{4}
\begin{picture}(50,15)(-25,0)
\put(0,0){\usebox{\dluu}}
\thicklines
\color{blue}
\qbezier(-2.5,9.5)(0,12)(2.5,14.5)
\qbezier(-2.5,14.5)(0,12)(2.5,9.5)
\qbezier(-2.5,3.5)(0,6)(2.5,8.5)
\qbezier(-2.5,8.5)(0,6)(2.5,3.5)
\color{red}
\qbezier(23,-2.5)(17,0)(11,2.5)
\end{picture}
-\,\,\frac{1}{4}
\begin{picture}(50,15)(-25,0)
\put(0,0){\usebox{\dluu}}
\thicklines
\color{blue}
\qbezier(-2.5,9.5)(0,12)(2.5,14.5)
\qbezier(-2.5,14.5)(0,12)(2.5,9.5)
\qbezier(-2.5,3.5)(0,6)(2.5,8.5)
\qbezier(-2.5,8.5)(0,6)(2.5,3.5)
\color{red}
\qbezier(-11,-2.5)(-17,0)(-23,2.5)
\end{picture}\quad.
\end{align}
From this result we observe another condition for the validity of~(\ref{eq:L_a_approx}),
namely that it should not matter whether the right
or the left vertex is differentiated with respect to time, i.e. the two vertices
should be equivalent. Whether this is the case, depends on the algebra of the model
under consideration. For noninteracting systems one can take bare vertices and the
reservoir indices of the two vertices have to be the same due to the reservoir 
contractions connecting them. In this case the condition is fulfilled if the
vertices do not depend on the level index of the dot states, e.g. through differently 
time-dependent coupling to different leads~\cite{Moskalets08}. For interacting systems,
the validity of~(\ref{eq:L_a_approx}) is more restrictive. The renormalized vertices
can be quite different from the bare ones and the vertices
get an additional dependence on the Laplace variable $E$ which is 
shifted by the chemical potentials of the reservoir lines 
crossing over the propagator standing left to that vertex. 
As a consequence, the two vertices are never equivalent
in the presence of a bias voltage and correction terms definitely occur for
time-dependent voltages. As discussed below, for the particular case of the IRLM
with one single reservoir, correction terms to eq.~(\ref{eq:L_a_approx}) do not appear in leading order.

\section{Results}
We use the above developed method to analyze the response of a mesoscopic capacitor
at zero temperature, described 
by the IRLM, where $H_{\rm res}=\sum_k \epsilon_k a^\dagger_k a_k$ describes a 
noninteracting reservoir with flat d.o.s. $\nu$ of band width $2D$,  
$H_{\rm dot}(t)=\epsilon(t) c^\dagger c$ denotes a spinless single-level 
quantum dot with time-dependent level position $\epsilon(t)$, and 
\begin{align}
\nonumber
V(t)&\quad=\quad\sqrt{\frac{\Gamma_0(t)}{2\pi\nu}}\,\sum_k\,(c^\dagger a_k+h.c.)\\
\label{eq:interaction}
&\hspace{0.8cm}
+\quad\frac{U(t)}{\nu}\,\sum_{kk'}\,(c^\dagger c-1/2)\,a^\dagger_k a_{k'}
\end{align}
is the dot-reservoir coupling with the bare time-dependent tunneling rate
$\Gamma_0(t)$ and the 
time-dependent dimensionless Coulomb interaction $U(t)$.

As shown above, we can evaluate the adiabatic response from eq.~(\ref{eq:rho_adiab}),
where $\Gamma^{(a)}$ and $\Gamma^{\prime (a)}$ can be extracted from eqs.~(\ref{eq:LG_final}) 
and~(\ref{eq:LU_final}) together with the RTRG results for the instantaneous vertices and the
Liouvillian derived in Ref.~\cite{rtrg_irlm}. For $E=0$ and leading order in $U$, 
the results of Ref.~\cite{rtrg_irlm} read
\begin{align}
\label{eq:gamma}
\Gamma&=\Gamma_0\left(\frac{D}{|\epsilon-i\Gamma/2|}\right)^{2U} \,,&
\partial_E\Gamma&=i\frac{U\Gamma^2}{\epsilon^2 + (\frac{\Gamma}{2})^2}\,,\\
\label{eq:gamma'}
\Gamma'&=-\frac{\Gamma}{\pi}\arctan \frac{\epsilon}{\Gamma/2} \,\,,&
\partial_E\Gamma'&=-\frac{i}{\pi}\frac{\Gamma\epsilon}{\epsilon^2+(\frac{\Gamma}{2})^2}\,,
\end{align}
where we have omitted the index $(i)$ for the instantaneous quantities.
Furthermore, the analysis in Ref.~\cite{rtrg_irlm} shows that the Coulomb vertex is zero 
in leading order for the Liouvillian elements containing $\Gamma$ and $\Gamma'$. Therefore  eq.~(\ref{eq:LG_final}) is sufficient  to evaluate 
$\Gamma^{(a)}$ and $\Gamma^{\prime (a)}$. Inserting
the algebra for the instantaneous tunneling vertices into eq.~(\ref{eq:LG_final}), one finds that eq.~(\ref{eq:L_a_approx}) is valid for the calculation
of $\Gamma^{\prime (a)}$, whereas for $\Gamma^{(a)}$ a correction term occurs 
proportional to $\partial_{t} U$. This yields the total result
\begin{align}
\label{eq:gamma_a}
\Gamma^{(a)}&=-\frac{U}{2}\partial_t\frac{\Gamma^2}{\epsilon^2+(\frac{\Gamma}{2})^2} \,,&
\Gamma^{\prime(a)}&=\frac{1}{2\pi}\partial_t\frac{\Gamma\epsilon}{\epsilon^2+(\frac{\Gamma}{2})^2} \,.
\end{align}
Since $\partial_E\Gamma,\Gamma^{(a)}\sim O(U)$ we can neglect them in leading order in  eq.~(\ref{eq:rho_adiab})
and, by inserting~(\ref{eq:gamma}) to~(\ref{eq:gamma_a}) into~(\ref{eq:rho_adiab}),
we find after a straightforward analysis for the charge response given by $Q=e\rho_+$,
\begin{equation}
\dot{Q}^{(i)}=C_0 \Gamma \partial_t \frac{\epsilon}{e\Gamma},\qquad
Q^{(a)}=-R_0 C_0^2 \Gamma \partial_t \frac{\epsilon}{e\Gamma},
\label{eq:response}
\end{equation}
where $R_0=\frac{h}{2e^2}$ and $C_0=\frac{e^2}{2\pi}\frac{\Gamma}{\epsilon^2+(\Gamma/2)^2}$.
In the special case of linear response and when only $\epsilon$ is varied with time, 
$C=C_0$ is the static capacitance and $R=R_0$ the universal relaxation resistance, in 
agreement with~(\ref{eq:RC}).
In contrast, when $\Gamma$ is varied with intent or via an accidental 
(but experimentally unavoidable) gate voltage dependence of $\Gamma_0$ or $U$, we 
obtain in linear response eq.~(\ref{eq:RC}) with
\begin{equation}
C=C_0\left(1-\frac{\epsilon}{\Gamma}\frac{\partial\Gamma}{\partial\epsilon}\right),\qquad
R=\frac{R_0 C_0}{C},
\end{equation}
where $\frac{\partial\Gamma}{\partial\epsilon}\approx 
\frac{\Gamma}{\Gamma_0}\frac{\partial\Gamma_0}{\partial\epsilon}+ 
2\Gamma\frac{\partial U}{\partial\epsilon}\log\frac{D}{|\epsilon-i\Gamma/2|}$.
As a consequence, $C$ and $R$ are very sensitive to the variation of other 
parameters, and logarithmic terms due to renormalization effects occur, if the Coulomb 
interaction $U$ varies with time. 

In this general case, where also the renormalized $\Gamma$ varies with time, we propose to analyze the time scale $\tau_Q$.
From~(\ref{eq:delay_time}) and~(\ref{eq:response})  we get
\begin{equation}
\tau_Q = \left|\frac{Q^{(a)}}{\dot{Q}^{(i)}}\right|=\frac{\Gamma/2 }{ \epsilon^2 + (\Gamma/2)^2} = R_0 C_0 \quad,
\label{eq:tau_Q}
\end{equation}
which is of $O(\Gamma^{-1})$ close to resonance $\epsilon\sim\Gamma$ and of 
$O(\Gamma/\epsilon^2)$ away from resonance. This result holds for {\it any} 
variation of $\epsilon$, $\Gamma_0$ and $U$ and is also valid in nonlinear response.
Interaction effects enter only weakly via the renormalized $\Gamma$ given by~(\ref{eq:gamma}).
Importantly, $\tau_Q$ reveals the static capacitance $C_0$ for a pure change of the 
gate voltage in linear response, with the advantage that $\tau_Q$ can be determined in the presence 
of the variation of any parameter. 

The experimentally accessible time scale $\tau_Q$ 
is thus an interesting quantity, which, for the case of the IRLM,  is stable for the
variation of any parameter in linear or nonlinear response. We note that this time 
scale can vary quite drastically if other observables or other models are studied.
E.g., the time scales $\tau_{Q}$ and $\tau_I$, when $Q$ is replaced by the current 
$I=\dot{Q}$, are in general the same only in linear response.
For the IRLM, the time scale $\tau_{I}$ shows similar 
logarithmic renormalizations in nonlinear response as they occur in $C$ and $R$ for time varying $U$.

\section{Realizations}
Several experimental realization of the IRLM exist, where the different parameters can be 
modulated in a controlled way.
As we outline here, the applicability of the IRLM is not limited to the description 
of an interacting quantum dot, but allows the observation of the predicted effects
for various physical systems.
First, we show that the low-energy physics of the IRLM is equivalent to the one of 
the Anderson-Holstein model, as first predicted in Ref.~\cite{schiller1}.
This model is widely used in molecular electronics~\cite{schoeller:vib} and describes a 
single-level molecular quantum dot, having a vibrational degree of freedom with frequency 
$\Omega$ coupled linearly to the charge of the dot
\begin{eqnarray}
H_\mathrm{dot} &=& \epsilon_M c^\dagger c + \Omega b^\dagger b  - \lambda\Omega (b+b^\dagger)c^\dagger c \,,\\
V &=& \sqrt{\frac{\Gamma_M}{ 2\pi\nu}}\sum_k(c^\dagger a_k+h.c.)\,.
\end{eqnarray}
The parameters $\epsilon_M$, $\Gamma_M$, $\lambda$ and $\Omega$ can be related to the 
effective parameters $\epsilon$, $\Gamma_0$ and $U$ of the IRLM.
Applying a Lang-Firsov transformation~\cite{langfirsov}, the coupling to the 
vibrational mode can be incorporated into the tunneling, leading to an effective 
level position, $\epsilon=\epsilon_M-\lambda^{2}\Omega$, and tunneling 
rate, $\Gamma_0= \Gamma_M e^{-\lambda^2}$~\cite{langfirsov,koch-lf}.
If the vibration frequency $\Omega$ is large compared to the other energy scales, the 
virtual intermediate states between the tunneling sequences with one or more bosons 
can be integrated out.
This produces terms with $n\geqslant 2$ lead operators in the effective Hamiltonian.
At large $\lambda$, all cotunneling processes with $n>2$ are 
exponentially suppressed, while the two-particle processes enter as an effective interaction.
Hence, by integrating out all vibrational modes the Anderson-Holstein model can be 
mapped onto the IRLM with effective interaction, $U=\frac{\Gamma_M}{2\pi\lambda^2 \Omega}$, 
with $\Omega\sim D$.
This is in agreement with Ref.~\cite{schiller1}, where it was shown numerically that this 
formula has even a broader range of applicability.
A modulation of the tunneling barriers is always accompanied by a modulation of the 
effective interaction $U$, since it is proportional to the tunneling rate $\Gamma_M$. 
In this case, our results predict that logarithmic corrections 
appear for $C$ and $R$, whereas the time scale $\tau_Q$ only depends on 
$\epsilon$ and $\Gamma$ via eqs.~(\ref{eq:tau_Q}) and~(\ref{eq:gamma}).
The Holstein coupling in the Hamiltonian allows for the 
observation of the dot charge via the displacement of the dot 
$\sim\langle b+b^{\dagger}\rangle$.

Finally, our results can be used to extract information on the relaxation behavior 
of systems described by the spin-boson model, namely, two-level dissipative systems 
connected to a large ensemble of oscillators
\begin{equation}
H= \frac{\epsilon}{2}\sigma_z-\frac{\Delta}{2}\sigma_x+\sum_q \omega_q b_q^\dagger b_q + 
\frac{\sigma_z}{2}\sum_q g_q (b_q+b_q^\dagger).
\end{equation}
The spin-boson model can be implemented by a Bose condensate of atoms trapped by a focused 
laser beam~\cite{recati}.
Such ultracold gases in optical lattices provide experimental realizations for theoretical 
models with remarkably independent tunability of parameters including the interaction 
strength, in contrast to usual semiconductor quantum dot setups.
The system's behavior depends crucially on the spectral coupling function. For the ohmic case, 
i.e. when the coupling constant obeys 
$\sum_q g_q^2\delta(\omega-\omega_q) =  2\alpha \omega e^{-\omega/D}$, 
the spin-boson model can be mapped onto the IRLM if $\alpha\approx 1/2$ 
(close to the Toulouse limit)~\cite{tsvelick-sbm,zwerger-sbm}.
The effective  IRLM parameters are
$U=1-\sqrt{2\alpha}$ and  $\Gamma_0=\Delta^2/D$.
Changing the coupling of the Bose condensate to the spin via $\alpha$ one 
generates a time-dependent effective interaction $U$.
The resulting response $\langle S_z\rangle$ of the spin, identified with $(\rho_+-\rho_-)/2$ in the effective IRLM, allows for the 
determination of the interesting time scale $\tau_{S_z}$, given by~(\ref{eq:tau_Q}).
Especially in the biased case, where $\tau_{S_z}\sim\Gamma/\epsilon^2$, this time scale is expected to differ significantly 
from typical relaxation rates $\Gamma$ and 
$\Gamma/2$ for the diagonal and nondiagonal components of the density matrix 
\cite{weiss,rtrg_irlm}.

\section{Conclusions}
In this Letter we provide a generic relation of the adiabatic response to
real-time RG results for the stationary case. The presented scheme allows 
for the variation of any parameter in linear or nonlinear response 
and provides criteria when the adiabatic correction to the Liouvillian can be 
calculated directly via energy and time derivatives of the instantaneous one. 
We suggest a delay time as an interesting time scale and show for the IRLM 
that its expression is robust against the choice of time-dependent parameters and their amplitude.
We confirm the universality of the AC relaxation resistance, unless a time dependence of tunneling and interaction is present, revealing logarithmic 
renormalizations due to charge fluctuations. 
We proposed different setups in molecular electronics and cold-atom systems to 
observe the effects experimentally.

\acknowledgements
We acknowledge valuable discussion with S. Andergassen and M. Pletyukhov, and 
useful comments by M. B\"uttiker, as well as financial support from the Ministry 
of Innovation NRW and DFG-FG 723.

\end{document}